\documentclass{elsarticle}
\usepackage{graphicx}
\usepackage{txfonts}

\def\dd{\mathrm{d}}

\usepackage{amssymb}

\usepackage{makeidx}

\usepackage{bbold} 

\def\simge{\mathrel{%
    \rlap{\raise 0.511ex \hbox{$>$}}{\lower 0.511ex \hbox{$\sim$}}}}
\def\simle{\mathrel{
    \rlap{\raise 0.511ex \hbox{$<$}}{\lower 0.511ex \hbox{$\sim$}}}}

\newcommand \be{\begin{eqnarray}}
\newcommand \ee{\end{eqnarray}}
\newcommand{\del}{\partial}

\journal{Physics Letters B}
\begin{document}


\begin{frontmatter}

\title{Burgers-like equation for spontaneous breakdown of the chiral symmetry in QCD}

\author{Jean-Paul Blaizot$^1$, Maciej A. Nowak$^2$, Piotr Warcho\l$^2$}

\address{$^1$Institut de Physique Th\'eorique, CNRS/URA2306, \\CEA-Saclay, 91191 Gif-sur-Yvette, France \\
 $^2$M. Smoluchowski Institute
of Physics and Mark Kac Center for Complex Systems Research,\\
Jagiellonian University, PL--30--059 Cracow, Poland}


\begin{abstract}
\noindent
We link the spontaneous breakdown of chiral symmetry in Euclidean QCD to the collision of spectral shock waves in the vicinity of zero eigenvalue of Dirac operator. The mechanism, originating from complex  Burger's-like equation for viscid, pressureless, one-dimensional flow of eigenvalues, is similar to recently observed weak-strong coupling phase transition in large $N_c$ Yang-Mills theory. The spectral viscosity is proportional to  the inverse of the size of the random matrix that replaces the Dirac  operator in the universal (ergodic)  regime. We obtain  the exact scaling function and critical exponents of the chiral phase transition for the averaged characteristic polynomial for $N_c \ge3$ QCD. 
We reinterpret our results in terms of known properties of chiral random matrix models and lattice data. 
\end{abstract}
\end{frontmatter}


\section{Introduction}
It was a recently emphasized  that the Burgers equation could be used  to understand universal features  of the weak to strong-coupling transition in two-dimensional Yang-Mills theory with a large number of colors $N_c$~\cite{BN,NEUBBURG,NEUBPRL}. This  transition, first studied by  Durhuus and Olesen ~\cite{DUROLE},  can be pictured, in the language of the Burgers equation, as resulting from the collision of two spectral shock waves at the closure of the gap. The emergence of the Burgers equation in such kind of problems, and related questions in random matrix theory, seems generic, and can be simply understood by exploiting Dyson's original idea of matrix random walks. Then, after a proper rescaling of time that separates the fast motion of the eigenvalues caused by their mutual repulsion, from the diffusion that results from the random walks of the matrix elements, one can easily show, by using standard tools of statistical mechanics, that the average resolvent obeys indeed a Burgers equation (or its simple generalizations) in the limit of large size of the matrices. Here we shall obtain the Burgers equation from an exact equation, akin to a diffusion equation, satisfied by  the average characteristic polynomial. The average resolvent and average characteristic polynomial  are simply related in the large $N$ limit. However, the equation for the characteristic polynomial is exact for any matrix size, while no such equation seems to exist for the average resolvent. 

Returning to two-dimensional Yang-Mills theory, we note that   the average characteristic determinant for a Wilson loop of a given area  was indeed shown to obey  an exact diffusion equation for any number of colors $N_c$ (which indeed yields a Burgers equation in  the infinite number of colors limit). 
For finite number of colors, the  flow of the eigenvalues is viscid, with a negative spectral viscosity $v_s=\frac{1}{2N_c}$. It is the negative sign of the viscosity, or of the diffusion constant,  that causes the rapid spectral oscillations at the closure of the gap. This picture was confirmed by extensive numerical studies of full Yang-Mills theory in three dimensions and recently, in four dimensions~\cite{NEUBPRL,NARNEU0607}.  

In this letter,  we demonstrate that a similar mechanism, governed by similar viscous spectral Burgers-like equation,  is  responsible for the universal spectral oscillations of the spectrum of the Dirac operator in QCD that accompany the spontaneous breakdown of the  chiral symmetry~\cite{VERZAH,SHURVER}.
 In the case of the Durhuus-Olesen  transition, where the role of the time is played by the area of the Wilson loop, an explicit  construction of a random matrix model was provided by Janik and Wieczorek~\cite{JANIKWIECZ}, matrices attached to large loops being formed by multiplying random matrices representing small loops. Here, as already mentioned,  following Dyson~\cite{DYSON}, we add a fictitious time (somewhat analogous to Schwinger's proper time) in order to describe the diffusion of the random matrices in  4 dimensional Euclidean space. 

\section{Random matrix theory and QCD}
A cornerstone for the microscopic understanding of the spontaneous breakdown of chiral symmetry  is the Banks-Casher~\cite{BC} relation
\be
|<\overline{q}q>|\equiv \Sigma =\pi \frac{\rho(0)}{V_4},
\label{bankscasher}  
\ee
where the quark condensate $<\overline{q} q>$ is an order parameter for chiral symmetry,  $\rho(0)$ is the averaged (over the gauge field configurations) level density of the Euclidean Dirac operator near the vanishing eigenvalue, and $V_4\equiv L^4$ is the Euclidean volume. This relation shows that chiral symmetry breaking requires a  strong accumulation of eigenvalues near zero, i.e. a level spacing $\Delta\sim 1/L^4$, much larger than the level spacing $\Delta\sim 1/L$ of a free system~\cite{LEUTSMIL}. This accumulation of eigenvalues leads to universal properties, that are well captured by random matrix theory:  For eigenvalues smaller than a characteristic energy scale, referred to as the Thouless scale $E_{Th}$,  the  fluctuations of eigenvalues are described by chiral random matrix models respecting the global symmetries of the Dirac Hamiltonian. In QCD,  the condition ${E_{Th}^{QCD}}/{\Delta}=F_{\pi}^2 L^2\gg1$, where $F_{\pi}$ is the pion decay constant,
determines the regime of applicability of random matrix theory  \cite{NJPZPRL,OSBORN}.

In Euclidean QCD, all four Dirac matrices can be chosen to be  anti-hermitian, hence the spectrum of the massless Dirac operator $D\equiv i\gamma_\mu (\del_\mu-igA_\mu)$ is purely imaginary. The partition function, for a fixed topological sector,  reads
\be
Z^{QCD}_{\nu}= \left< \prod_f m_f^{\nu} \prod_l (\lambda_l^2 +m_f^2)   \right>_\nu
\label{ZQCD}
\ee
where the averaging is done with respect to gluonic configurations of a given topological  charge $\nu$,  $\pm i\lambda_k$ are the eigenvalues of $D$,   and $m_f$ is the mass of a  quark with flavor $f$.  Due to the chiral symmetry non-zero eigenvalues of $D$ come in pairs, and the number of fermionic zero modes is related  to the topological charge.   
In the chiral Gaussian random matrix model (hereafter $\chi$GUE), corresponding to QCD with $N_c \ge 3$, the role of the massless Dirac operator is played by a random matrix 
$W (= -iD)$ of the following form:
\be
W=\left(\begin{array}{cc} 0 & K^{\dagger} \\ K & 0 \end{array}\right).\label{chiral1}
\ee
Here  $K$ is a rectangular $M\times N$ ($M>N$) matrix with complex entries,  $K_{ij}\equiv x_{ij}+iy_{ij}$, where $x_{ij}$ and $y_{ij}$ are drawn from a  Gaussian   
distribution. Note that $W$ is hermitian, so that its eigenvalues $\kappa_i$'s are real. The block-diagonal structure of (\ref{chiral1}) reflects  the chiral symmetry of the Dirac operator: $W$ anticommutes with the analogue of the Dirac matrix $\gamma_5$, defined  here as $\gamma_5={\rm diag}({\bf 1}_N, -{\bf 1}_M)$. This implies in particular that the eigenvalues come in pairs of opposite values, 
($\kappa,-\kappa$). 
By construction, $W$ has in addition $\nu\equiv M-N$ zero eigenvalues. These mimic the zero modes of quarks propagating in gauge fields of non trivial topology. 

General spectral properties of the random matrices can be obtained form  correlation functions 
containing both products and ratios of the characteristic polynomial~\cite{FYOSTR}
\be C(w_1,....w_f;w^{'}_1,....w^{'}_b) =\left<   \frac{\prod_{i=1}^f Z(w_i)}{\prod_{j=1}^b Z(w^{'}_j)}   \right>  
\label{korellator}
\ee
where $<...>$ denote averaging with respect to the ${\chi}$GUE measure, and the characteristic polynomial is $ Z(w)=\det(w-W)$.
Directly related to (\ref{korellator})  is the resolvent, $R(z)=\partial_w C(z;w)|_{w=z}$, whose imaginary part yields the spectral density. 
 In the vicinity of zero, the microscopic (unfolded) resolvent  predicts for QCD~\cite{VEROXF} 
\be\label{averageresolvent}
\frac{R(z)}{V_4 \Sigma}= x\, (I_b(x) K_b(x) +I_{b+1}K_{b-1})
\ee
where $b=N_f +|\nu|$, with $N_f$ the number of quark flavors,  and $x=zV_4 \Sigma$. The appearance of Bessel functions $I_b$, $K_b$ in the correlation functions, such as in Eq.~(\ref{averageresolvent}), is generic. They encode the universal  behaviors that show up  already in the simplest objects, the average characteristic polynomial and/or the average of the inverse of the characteristic polynomial. As the first result of the paper, we shall obtain an exact differential equation for the averaged characteristic polynomial. This equation is akin to a diffusion equation, from which a Burgers equation can be derived from a simple transformation.

 \section{Burgers equation and the characteristic polynomial for Dirac operator}
We assume now  that the entries of the matrix $K$  
 follow independent random walks.   Let us denote by  $P(X,Y,t)$ the joint probability that the entries  of $K$ take the values $X=\{x_{ij}\}$ and $Y=\{  y_{ij}\}$ at time $t$. The random walks of the matrix elements translates into the following diffusion equation for  $P(X,Y,t)$:
\be
\del_{t}P\left(X,Y,t\right)=\frac{1}{4}\sum_{i=1}^M\sum_{j=1}^N(\del^2_{X}+\del^2_Y)P\left(X,Y,t\right)\label{diff}.
\ee

We shall be interested in this paper with the time evolution of the averaged characteristic polynomial
\be
	Q^{\nu}_{n}(w,t)\equiv\left\langle{{\rm det}{\left[w-W\right]}}\right\rangle
= \int {\rm d}X\,{\rm d}Y\, P(X,Y,t)\, \det[w-W(X,Y)],	
	\label{characteristic1}
\ee	
where $n=N+M$, $w$ is an arbitrary complex number, and ${\rm d}X=\prod_{ij} {\rm d}x_{ij}$, and similarly for ${\rm d}Y$. In order to get the equation obeyed by $Q^{\nu}_{n}(w,t)$, we  consider  first the equation satisfied by the average characteristic polynomial of the associated $N\times N$ Wishart matrix $K^\dagger K$, $M^{\nu}_N\equiv\left\langle{{\rm det}{\left[z-K^{\dagger}K\right]}}\right\rangle$. This,  we write as an integral over $N$  Grassmann variables  $\eta_i,\overline\eta_i$:
\be
M^{\nu}_N	(z, t)=\int{\rm d}\eta \,{\rm d}\overline{\eta }\, {\rm e}^{z\sum_i{\overline{\eta }}_i\eta_i  }\,\int{\rm d}X\dd {Y} \,P\left(X,Y,t\right)\,\exp{\left({-\overline{\eta}K^{\dagger}K\eta }\right)},
	\ee
	where $\overline{\eta}K^{\dagger}K\eta =\sum_{i,k=1}^N\sum_{j=1}^M \overline{\eta}_i (x_{ij}-iy_{ij})(x_{jk}+iy_{jk})\eta_k$.
To derive the equation for $M$, we take a time derivative of the expression above. This acts on $P(X,Y,t)$, which, using   Eq.~(\ref{diff}), we transform into derivatives with respect to $x_{ij}$ and $y_{ij}$. Then, we  integrate by parts, and use standard  Grassmann calculus to  obtain (after a somewhat tedious but straightforward calculation) the following differential equation 
\be
\del_{t}M_{N}^{\nu}(z,t)=-z\del^2_{z}M_{N}^{\nu}(z,t)-(\nu+1)\,\del_{z}M_{N}^{\nu}(z,t)\label{pde1}.
\ee
This equation is valid for any $N$ and $M$, and arbitrary initial conditions. Note that  for the trivial initial condition $K_{ij}(t=0)=0$, its solution is given  by time dependent associated Laguerre polynomial \cite{BNW1}.

 From the equation for $M_{N}^{\nu}(z,t)$, Eq.~(\ref{pde1}) above, one easily obtains the equation for $Q_{N}^{\nu}(w,t)=w^\nu\,M_{N}^{\nu}(z=w^2,t)$.  This   reads
\be
\del_{t}Q_{n}^{\nu}(w,t)=-\frac{1}{4}\del^2_{w}Q_{n}^{\nu}(w,t)-\frac{1}{4w}\del_{w}Q_{n}^{\nu}(w,t)+\frac{\nu^{2}}{4w^2}Q_{n}^{\nu}(w,t).
\label{maindiffeq}
\ee
It will be also useful to consider the equation for the Cole-Hopf transform of $Q_{n}^{\nu}(w,t)$,  $f^\nu_{n}\equiv \frac{1}{n}\del_{w}{\rm ln}(Q^{\nu}_{n}(w,t))$. This object identifies with the average resolvent in the large $n$ limit.  After a rescaling of the time,  $\tau=M t$, one gets from Eq.~(\ref{maindiffeq})
\be
\frac{n+\nu}{n}\del_{\tau}f_{n}^\nu+f_{n}^\nu\,\del_{w}f_{n}^\nu=-\frac{1}{2n}\left[  \del^2_{w}f_{n}^\nu+\frac{1}{w}\del_{w}f_{n}^\nu-\frac{1}{w^{2}}f_{n}^\nu\right]-\left(\frac{\nu}{n}\right)^{2}\frac{1}{w^3 },\label{chiralr}
\ee
where we have separated on  the left hand side the terms that survive the large $n$ limit, and on the right hand side the terms that are explicitly  suppressed by powers of $1/n$. Note the crucial role played by the rescaling of time in arriving at this equation. The motivation behind this rescaling is that the diffusion associated with the random walks is taking place over a time scale that is  larger, typically by a factor $n$, than the time scale corresponding to the local rearrangements of the eigenvalues due to their mutual repulsion \cite{DYSON,BNW1}.
After rescaling, the diffusion terms are dwarfed by a factor $1/n$, and the large $n$ dynamics is dominated by repulsion. The last term, of order $1/n^2$ finds it origin in the kinematical zero modes present when $\nu\ne 0$.

\section{Large n limit}

We consider now the limit  $n\to\infty$, with $\nu$ constant. We set $g(w,\tau)=\lim_{n\to\infty}f_{n}^\nu(w,\tau)$. Eq.~(\ref{chiralr}) reduces then to the inviscid Burgers equation \cite{BN,BURGERS,ZAKOPANE},  independent of $\nu$:
\be
\del_{\tau}g(w,\tau)+g(w,\tau)\del_{w}g(w,\tau)=0\label{chig}.
\ee
It  can be solved using complex characteristics. 
We  choose the system to be   initially in a chiral symmetric state, with eigenvalues localized at $\pm a$.The characteristic lines are given by $w=\xi+\tau g_0(\xi)$, with 
\be
g_{0}(w)\equiv g(w,\tau=0)=\frac{1}{2}\left(\frac{1}{w-a}+\frac{1}{w+a}\right)=\frac{w}{w^{2}-a^{2}}.
\label{cubic}
\ee
The solution $g(w,\tau)$ is constant along the characteristics, meaning $g(w,\tau)=g_{0}\left[\xi(w,\tau)\right]$. By eliminating $\xi$, one obtains an implicit equation for $g$:
\be
\tau^2 g^3 -2\tau w g^2 +(w^2 -a^2 +\tau)g -w=0 \label{chig1}.
\ee
This equation can be solved by elementary means, and well known results recovered. In fact, the change of variables $\omega=w/\sqrt{\tau}, d=a/\sqrt{\tau}$, and $G_{st}(\omega)=\sqrt{\tau}\,g(\omega\sqrt{\tau},\tau)$ transforms this equation into $G_{st}^3-2\omega G_{st}^2+(\omega^2-d^2+1)G_{st}=\omega$, an equation for a time-independent resolvent $G_{st}$ that has been obtained in this context using different techniques \cite{JACVER,BLUE,WEIDEN}. In  previous studies, the parameter $d$ was introduced as the ``deterministic" part of the chiral matrix (with $K$ in  Eq.~(\ref{chiral1}) replaced by $K+d$,  $d$ being fixed and $K$ random), in order to control the approach to the chiral transition. The aforementioned change of variables renders transparent the dynamics captured by the Burgers equation:  For  small time, i.e. $\tau<a^2$,  the spectral density remains localized in humps  centered around the values $\pm a$. As time reaches the critical value $\tau_c=a^2$ (corresponding to $d=1$ in the static approach), the two domains of the spectrum merge at the origin, which we picture as the collision of two spectral shock waves. A finite ``condensate'' then develops,  and chiral symmetry is spontaneously broken.

We now recover these features by studying the singularities of the characteristics, and the behavior of the solution in the vicinity of these singularities. This brief discussion will pave the way for the scaling analysis to be performed in the next section. Singularities appear when characteristics start to cross (appearance of a spectral shock wave).  This occurs for values of $\xi$ that obey the equation, 
\be
\left.\frac{\dd w}{\dd \xi}\right|_{\xi=\xi_c} =0=1+\tau g_0'(\xi_c),
\ee
that is 
\be
\tau \left(a^2+\xi_{c}^2\right) =\left(\xi_{c}^{2}-a^2 \right)^2.
\ee
The values of  $\xi_c$ correspond to the edges of the spectrum when $\tau<\tau_c=a^2$. 
At the critical time $\tau_c=a^2$, where  the two humps of the spectrum start to merge,  the equation for $\xi_c$ admits a double solution at $\xi_c=0$, which splits into two purely imaginary and opposite solutions when $\tau>a^2$.  

In order to study the behavior of $g$ at the edge of the spectrum, we expand $g_0$ in the vicinity of a singular point
\be
g_{0}(\xi)=g_{0}(\xi_c )+(\xi-\xi_c )g_{0}'(\xi_c )+\frac{1}{2}(\xi-\xi_c )^{2}g_{0}''(\xi_c )+\frac{1}{6}(\xi-\xi_c )g_{0}'''(\xi_c )+\ldots
\ee
When $\tau<a^2$, $g_{0}'(\xi_c )=-{1}/{\tau}$ and, using $g_0=(w-\xi)/\tau$, we get 
\be
w-w_c =\frac{\tau}{2}(\xi-\xi_c )^{2}g_{0}''(\xi_c )+\ldots
\ee
One can then easily invert the relation between $w$ and $\xi$, and get (for the rightmost edge)
 \be
 \xi-\xi_c=\pm\sqrt{\frac{2}{\tau g_0''(\xi_c)}}\sqrt{w-w_c},
 \ee
so that
 \be\label{gwtausr}
g(w,\tau)=g_{0}(\xi)\simeq g_{0}(\xi_c )+(\xi-\xi_c )g_{0}'(\xi_c )= g_{0}(\xi_c )\mp\frac{1}{\tau}\sqrt{\frac{2}{\tau g_0''(\xi_c)}}\sqrt{w-w_c},
\ee
which exhibits  the familiar square root behavior of the spectrum  near its (right) edge. 
 
 For $\tau=a^2$, a similar analysis taking into account that $g_{0}(\xi_c )=0=g_{0}''(\xi_c )$, so that the cubic term must be kept, yields (for $w_c=0$)
\be
g(w,\tau)=\left(-\frac{w}{a^2}   \right)^{1/3}.
\ee

For time $\tau>a^2$,  we have $g(w=0,\tau)=g_{0}\left(\xi_{c}(w=0,\tau)\right)=-\frac{\sqrt{a^2 -\tau}}{\tau}$,  which is imaginary and hence directly proportional to  the spectral density $\rho(0)$.

The  behavior of $g(w)$ at the edge of the spectrum determines the average eigenvalue spacing in the limit of large matrices. A singularity  $\sim|w-w_c|^\alpha$ yields a level spacing $\sim n^{-\delta}$ with $\delta=1/(1+\alpha)$. We have therefore $\delta=2/3$ for $\tau<a^2$, $\delta =3/4$ for $\tau=a^2$ and $\delta =1$ for $\tau>a^2$. We shall exploit these properties in the next section.

\section{Critical properties of the characteristic polynomial}

In this section we  carry out a scaling analysis of the average characteristic polynomial $Q(w,\tau)$, or its Cole-Hopf transform $f(w,\tau)$, in the vicinity of the singular points. To that aim, we  set 
\be
w=w_c+n^{-\delta} s, \qquad  f^\nu_{N}\to g_{0}(\xi_{c})+n^{-\gamma}\chi(s,\tau),
\ee
 with $\gamma=1-\delta$, and  $s$ and $\chi(s,\tau)$ remain finite  as $n\to\infty$. 

\subsubsection{Airy edge}

Let us focus first  on the left  edge of the positive part of the spectrum, for $\tau<a^2$, where we expect the solution to be of the form (see Eq.~(\ref{gwtausr}))
\be
f^{\nu}_{n}(w,\tau)\approx g_{0}(\xi_{c})+g'_{0}(\xi_{c})\sqrt{\frac{2(w-w_{c})}{\tau g''_{0}(\xi_{c})}}.
\ee
This exhibits a square root singularity ($\delta=2/3$), and in line with the discussion above we set
  $
w=w_{c}+n^{-\frac{2}{3}}s$ and $ f^\nu_{n}\to g_{0}(\xi_{c})+n^{-\frac{1}{3}}\chi(s,\tau).
$
(Notice that $g_0(\xi_c) $ is a function of $\tau$ only, $g_{0}(\xi_{c})=\del_{\tau}w_{c}$.)
Substituting this ansatz into Eq.~(\ref{chiralr}),  and keeping only the dominant terms as $n\to\infty$,  we get (with $\dot g_0(\xi_c)=\del_\tau g_0(\xi_c)$)
 \be
\dot{g}_{0}(\xi_{c})+\chi\,\del_{s}\chi+\frac{1}{2}\del^2_{s}\chi=0= \del_{p}\left(2p\dot{g}_{0}(\xi_{c})+\chi^{2}+\del_{s}\chi\right).
\ee
This equation is easily integrated, $\chi^{2}+\del_{s}\chi+2s\dot{g}_{0}(\xi_{c})+u(\tau)=0$, 
with $u(\tau)$ an arbitrary function of $\tau$. 
Defining $\chi(s,\tau)=\del_{s}\ln\phi(s,\tau)$ and shifting $s=\tilde{s}-{u(\tau)}/({2\dot{g}_{0}(\xi_{c})})$ (with $\phi(s,\tau)=\psi(\tilde{s},\tau)$) one transforms the equation above into the equation for the Airy function, 
$
\del^2_{\tilde{p}}\psi+2\dot{g}_{0}(\xi_{c})\tilde{p}\psi=0.
$
The looked for solution is therefore 
\be
\phi(s,\tau)=Ai\left[-(2\dot{g}_{0})^{\frac{1}{3}}(s+\frac{u}{2\dot{g}_{0}})\right].
\ee
The arbitrary function $u(\tau)$ can be determined by a careful matching of the asymptotic form as $s\to\infty$ of the solution to its large $N$ limit. One then finds that $u$ actually vanishes~\cite{BNW1}.

\subsubsection{Bessel universality}
For $\tau > a^2$, there is no singular behavior ($\delta=1$). So, we  set  $w=n^{-1}s$ and $f^{\nu}_{n}=\chi$. In the large $n$ limit (at $\nu$ constant) we obtain, following the same manipulations as above,   the following partial differential equation:
\be
0=\del_{s}\left(\chi^{2}+\del_{s}\chi+\frac{\chi}{s}-\frac{\nu^{2}}{s^{2}}\right),
\label{Besselinitial}
\ee
which integrates to 
$\chi^{2}+\del_{s}\chi+\frac{\chi}{s}-\frac{\nu^{2}}{s^{2}}+u(\tau)=0.
$
 Then, setting  $\chi(s,\tau)=\del_{s}\ln\phi(s,\tau)$ we obtain:
\be
s^{2}\del^2_{s}\phi+s\del_{s}\phi+\phi\left[s^2 u(\tau)-\nu^{2}\right]=0,
\ee
whose solution is
\be
\phi(s,\tau)= J_{\nu}\left[s\sqrt{u(\tau)}\right].
\ee
The determination of the arbitrary function $u(\tau)$ proceeds as in the previous case, by matching the asymptotic $\chi(s,\tau) \sim -i\sqrt{u(\tau)}$ with the large $N$ solution. This gives $\sqrt{u(\tau)}=\sqrt{\tau-a^2}/\tau.$  We recover the scaling of the  ratio of spectral densities discussed for instance in \cite{SENER,WETTIG,DAMGAARD}.

\subsubsection{Bessoid (axially symmetric Pearcey) universality}

Finally, we move to the case of $\tau=\tau_c=a^2$, which will lead to the second new result of this paper.  As we have shown above, at $\tau=a^2$, the two pre-shocks collide. Before the collision,  these pre-shocks are accompanied by oscillations of the Airy type. Our purpose now is to describe the modification of the pattern of oscillations for  $\tau$ close to $\tau_c$. To perform this analysis, it is most convenient to start from the diffusion equation obeyed by the characteristic polynomial, i.e., Eq.~(\ref{maindiffeq}) which, after changing $t$ into $\tau=M t$, we rewrite as 
\be
\del_{\tau}Q_{n}^{\nu}(w,\tau)=-\frac{1}{4M}\del^2_{w}Q_{n}^{\nu}(w,\tau)-\frac{1}{4Mw}\del_{w}Q_{n}^{\nu}(w,\tau)+\frac{\nu^{2}}{4Mw^{2}}Q_{n}^{\nu}(w,\tau).
\label{Polyanin}
\ee
This equation is to be solved with the initial condition $Q^{\nu}_{n}(w,\tau=0)=w^{\nu}\left(w^{2}-a^{2}\right)^{N}$ (where the function of $w$, $(w^2-a^2)^N$  is defined with a cut between $-a$ and $a$). 
It can be verified by a direct calculation (that does not require the explicit calculation of the integral below), that 
\be\label{integralrepres}
Q_{n}^{\nu}(w,\tau)= \mathcal{C} \,\tau^{-1} \int^{\exp({i\phi_{y}})\infty}_{0} y^{\nu +1} \exp{\left(M\frac{w^2+y^{2}}
{\tau}\right)}I_{\nu} \left(\frac{2M y w}{\tau}\right)(y^{2}-a^{2})^{N}\dd y,
\ee
where $\mathcal{C}=(-1)^{\nu+1/2}2 M $, is a solution with the proper initial condition.  (For $\nu=0$, this solution  agrees with a known solution~\cite{POL}.) The $y$-integral runs over a half-line that starts at the origin and goes to infinity, making a constant angle $\phi_y=\arg(y)$ with the real axis, with $-\pi\le \arg(y)<\pi$.
 For the integral to be convergent, we require $\frac{\pi}{4}<|\phi_y|<\frac{3\pi}{4}$. The modified Bessel function  $I_\nu(x)$ had the following asymptotic expansion $\lim_{|x|\to\infty}I_{\nu}(x)\simeq\frac{1}{\sqrt{2\pi x}}e^{x}$, valid for $|\arg(x)|<\frac{\pi}{2}$ (see \cite{ABRA}; here $x=\frac{2Myw}{\tau}$). This is useful in particular to verify the initial behavior. Indeed, as $\tau\to 0$, one may estimate the integral using the saddle point method. The saddle point equation yields $y+w=0$, which fixes in particular $\arg(y)=\arg(-w)$. This new condition for $\phi_{y}$, together with the convergence condition noted above, are easily seen to be compatible with the condition of validity of the asymptotic expansion of the Bessel function. In turn, these conditions limit the allowed arguments of $w$ to $\frac{\pi}{4}<|\arg(w)|<\frac{3\pi}{4}$.

The integral representation (\ref{integralrepres})  of the characteristic polynomial allows us to study  the vicinity of the critical point. We note that the saddle point equation reads (in the large $n$ limit)
\be
\frac{2y}{\tau}+\frac{2 w}{\tau}+\frac{y}{y^2-a^2}=0.
\ee
Identifying $y=-\xi$, we recognize the equation for the characteristic lines. This indicates how the large $n$ dynamics is coded in this integral. We shall focus more specifically at the critical point, $w=0, y=0, \tau=a^2$. In this regime, we may expand  $\ln(a^{2}-y^{2})\approx \ln(a^{2})-\frac{y^{2}}{a^{2}}-\frac{y^{4}}{2a^{4}}$, and obtain 
\be\label{bessoid1}
a^{2N}\mathcal{C}' \, \tau^{-1}  \exp{\left(\frac{M w^2}{\tau}\right)}\int^{\exp({i\phi_{y}})\infty}_{0}y^{\nu +1}\exp{\left[-\frac{Ny^{4}}{2a^{4}}-\frac{Ny^{2}}{a^{2}}+\frac{M y^{2}}{\tau}\right]}I_{\nu}\left(\frac{2Myw}{\tau}\right)\dd y,
\ee
with $ \mathcal{C}'=\mathcal{C}(-1)^N $.
To capture the critical behavior also as a function of time, we set $\tau=a^{2}+\theta$.  Then Eq.~(\ref{bessoid1}) becomes
\be
a^{2(N-1)} \mathcal{C}' \, \exp{\left(\frac{Mw^2}{a^{2}}\right)}\int^{\exp({i\phi_{y}})\infty}_{0}y^{\nu +1}\exp{\left[-\frac{Ny^{4}}{2a^{4}}+\nu\frac{y^{2}}{a^{2}}-\frac{My^{2} \theta}{a^{4}}\right]}I_{\nu}\left(\frac{2Myw}{a^{2}}\right)\dd y.
\ee
This expression suggests the following change of  variables that will ensure a smooth large $n$ limit:
\be
y= n^{-\frac{1}{4}}au,\,\,\,\,
w=  n^{-\frac{3}{4}}r a q,\,\,\,\,
t= n^{-\frac{1}{2}}ra^{2}\theta. \label{rescalings}
\ee
We then define 
\be
B^{\nu}(q,\theta)\equiv \lim_{N,M\to\infty}\mathcal{C'}^{-1} a^{-n} n^{\frac{\nu+2}{4}} (-1)^{\nu+1} Q_{n}^{\nu}\left(n^{-\frac{3}{4}}r a q,\,\,a^{2}+\frac{1}{2}n^{-\frac{1}{2}}ra^{2}\theta\right),
\ee
where the limit is taken with $\nu$ constant. Finally
\be\label{bessoid}
B^{\nu}(q,\theta)=(-1)^{\nu+1}\int_{0}^{\exp({i\phi_{q}\pm i\pi})\infty}u^{\nu +1}\exp{\left(-\frac{1}{4}u^{4}-\frac{1}{2}u^{2}\theta\right)}\,I_{\nu}\left(qu\right)\dd u.
\ee
This exact scaling function for the characteristic polynomial is the second important result of this paper. It has a form very similar to the Pearcey function, which gives  the asymptotic behavior of the characteristic polynomial in the case when gap closes for GUE~\cite{BREZINHIKAMI} or in the case of  unitary diffusion on 
the circle~\cite{NARNEU0607}  
\be
P(q,t)= \int_{-\infty}^{\infty} dy  \exp (-y^4 -t y^2 +qy).
\label{Pearcey}
\ee
Here $q$ is the rescaled  angle representing the position of the eigenvalue on the unitary circle and $t$ parameterizes the fluctuations around the critical area.  The critical indices that determine the scaling with $n$ in Eq.~(\ref{rescalings}) are identical to those in the Pearcey integral,  but the form of the integral is different. 
The reason is the chiral symmetry, which  imposes an additional polar symmetry of the spectrum, trading the exponential  function of $q$ in  the Pearcey integral (\ref{Pearcey}) for a Bessel function in Eq.~(\ref{bessoid}).

\section{Conclusions}
In this letter we have obtained an exact  differential equation for the average characteristic polynomial of a chiral random matrix, and its Cole-Hopf transform. For the latter, the equation takes the form of a generalized viscid Burgers equation, where the viscosity is proportional to the inverse of the size of the matrix, but with a negative sign. This allowed us to provide a complete description of the  full critical behavior of averaged characteristic polynomials in chiral QCD, based on a single equation.  In particular, we considered the case of chiral Gaussian Unitary Ensembles and we have identified the exact universal scaling function (Bessoid $B^{\nu}(q, \theta)$) in the vicinity of the chiral critical point for the average characteristic polynomial.  
We did not analyze in this letter the properties of the average of the inverse characteristic polynomial, but we have checked that if fulfills similar equations, albeit with initial conditions singular at $w=0$, alike in the cases of unitary and GUE diffusions. 

Since $F_{\pi}$ scales like $\sqrt{N_c}$, more and more eigenvalues of the Dirac operator fall into the universal window when the number of colors tends to infinity,  the volume of the lattice being kept finite. This suggests, that the present study is  relevant also for  analyzing the spontaneous breakdown of chiral symmetry at finite volume and large $N_c$ QCD, which  was  observed, and explained by Neuberger and Narayanan~\cite{NN1}. For small lattice sizes, chiral symmetry is unbroken, while at some critical scale $L_c$ a condensate is formed.
The same authors~\cite{NN2}  have also observed, that the $N_c$-dependence of the level spacing closest to zero goes from $1/N_c$ in the broken chiral symmetry phase to $1/N_c^{2/3}$ in the symmetric (gapped) phase. At
$L=L_c$, the critical scaling changes to $1/N^{3/4}$  behavior and the  condensate vanishes at the critical size $L_c$ as $\sqrt{L-L_c}$. These results are in agreement with our analysis.

The critical universal scaling function (Bessoid) for large $N_c$ Dirac operator resembles closely the critical universal  scaling  function (Pearcey's cuspoid) for the weak to strong coupling transition in Yang-Mills theory at large $N_c$. It would be interesting to study numerically both transitions simultaneously (at least in some simple model like \cite{NN1}) to see the interplay between the cuspoid and the Bessoid, or, in other words, the relation between the critical size $L_c$ for chiral symmetry breakdown and the critical area $\sim L_c^2$  for the weak to strong coupling transition in Yang-Mills theory. Of particular interest would be to measure on the lattice  the microscopic spectral density exactly at the point of the transition.  As far as we know, the microscopic spectral density at criticality was constructed explicitly only for $b=N_f+|\nu|=0$ case~\cite{CRITICALUS}, and  was never checked by the lattice simulation.  
Taking into account the still  ongoing discussion on the nature of chiral phase transition, its relation to 
confinement and  Anderson localization~\cite{GARCIA,RECENT},  lattice verification  of analytic predictions for microscopic  densities at the critical point may be a powerful  tool to shed more light on this aspect of strong interactions.

\section*{Acknowledgements}
PW is supported by the International PhD Projects Programme of the Foundation for Polish Science within the European Regional Development Fund of the European Union, agreement no MPD/2009/6. 
MAN is supported in part by the Grant DEC-2011/02/A/ST1/00119 of the  National Centre of Science.

\end{document}